\documentclass[12pt]{article}
\usepackage{graphics}
\begin{document}

\title{Entanglement of pair cat states and teleportation}
\author{Shang-Bin Li, Ru-Kuan Wu, Qin-Mei Wang, Jing-Bo Xu\\
Chinese Center of Advanced Science and Technology (World Labor-\\
atory), P.O.Box 8730, Beijing, People's Republic of China;\\
Zhejiang Institute of Modern Physics and Department of Physics,\\
Zhejiang University, Hangzhou 310027, People's Republic of
China\thanks{Mailing address}}
\date{}

\maketitle

\begin{abstract}
{\normalsize The entanglement of pair cat states in the phase
damping channel is studied by employing the relative entropy of
entanglement. It is shown that the pair cat states can always be
distillable in the phase damping channel. Furthermore, we analyze
the fidelity of teleportation for the pair cat states by using
joint measurements of the photon-number sum and phase difference.
\\

}

\end{abstract}
\newpage
\section * {I. INTRODUCTION}
\hspace*{8mm}Quantum entanglement plays an important role in
various fields of quantum information, such as quantum computation
[1], quantum cryptography [2], quantum teleportation [3,4], dense
coding [5] and quantum communication [6], etc. It has been
recognized [7,8] that quantum teleportation can be viewed as an
achievable experimental technique to quantitatively investigate
quantum entanglement. There exists a class of states called
maximally correlated states, which have an interesting property,
i.e., the PPT distillable entanglement is exactly the same as the
relative entropy of entanglement [9]. Both two-mode squeezed
vacuum states and pair cat states [10] belong to this class. In
continuous variable teleportation the entanglement resource is
usually the two-mode squeezed state, or the
Einstein-Podolsky-Rosen (EPR) states. Continuous variable quantum
teleportation of arbitrary coherent states has been realized
experimentally by employing a two-mode squeezed vacuum state as an
entanglement resource [7]. Theoretical proposals of teleportation
scheme based on the other continuous variable entangled states
have already been discussed [11]. Update now, little attention has
been paid to the entanglement properties of the pair cat state and
its possible application of quantum information. Gou et al. have
proposed a scheme for generating the pair cat state of motion in a
two-dimensional ion trap [12]. In their scheme, the trapped ion is
excited bichromatically by five laser beams along different
directions in the X-Y plane of the ion trap. Four of these have
the same frequency and can be derived from the same source,
reducing the demands on the experimentalist. It is shown that if
the initial vibrational state is given by a two-mode Fock state,
pair cat states are realized when the system reaches its steady
state. Their work motivates us to investigate the entanglement
properties of the pair cat state and its possible application in
quantum information processes, such as quantum teleportation. The
motivation is two-fold: (1) the storage of continuous variable
entangled states in two-dimensional motional states of trapped
ions is feasible in current experimental techniques. (2) the
mapping of steady state entanglement to optical beams is also
realizable [13].\\
\hspace*{8mm}On the other hand, quantum entanglement is a fragile
nature, which can be destroyed by the interaction between the real
quantum system and its environment. This effect, called
decoherence, is the most serious problem for all entanglement
manipulations in quantum information processing. There have
several proposals for entanglement distillation and purification
in continuous variable systems [14]. In this paper, we firstly
investigate the relative entropy of entanglement of pair cat
states in the phase damping channel, and show that the pair cat
states can always be distillable in the phase damping channel.
Then, we explore possible application of pair cat states in
quantum information processing, such as quantum teleportation. The
fidelity of teleportation protocol in which the mixed pair cat
state is used
as a entangled resource, is analyzed.\\
\hspace*{8mm}This paper is organized as follows. In section II,
based on the exact solution of the master equation describing
phase damping, we give the numerical calculations of relative
entropy of entanglement for pair cat states in the phase damping
channel and investigate the influence of the initial parameters of
these states on the relative entropy of entanglement. In section
III, we analyze the fidelity of teleportation for the pair cat
states by using joint measurements of the photon-number sum and
phase difference. The influence of phase damping on the fidelity
is discussed. A conclusion is
given in section IV.\\

\section * {II. RELATIVE ENTROPY OF ENTANGLEMENT OF PAIR CAT STATES IN PHASE DAMPING CHANNEL}
\hspace*{8mm}The relative entropy of entanglement is a good
measure of quantum entanglement, it reduces to the Von Neumann
entropy of the reduced density operator of either subsystems for
pure states. For a mixed state $\rho$, the relative entropy of
entanglement [15] is defined by
$E_R(\rho)=\min_{\sigma\in{D}}S(\rho\parallel\sigma)$, where $D$
is the set of all disentangled states, and
$S(\rho\parallel\sigma)=\textrm{Tr}[\rho(\log_2\rho-\log_2\sigma)]$
is the quantum relative entropy. It is usually hard to calculate
the relative entropy of entanglement for mixed states. Recently,
it has been shown [27] that the relative entropy of entanglement
for a class of mixed states characterized by the following density
matrix
$$
\rho=\sum_{n_1,n_2}a_{n_1,n_2}|\phi_{n_1},\psi_{n_1}\rangle\langle\phi_{n_2},\psi_{n_2}|
\eqno{(1)}
$$
can be written as
$$
E_R(\rho)=-\sum_{n}a_{n,n}\log_{2}a_{n,n}+\textrm{Tr}(\rho\log_2\rho).
\eqno{(2)}
$$
The separate state $\rho^{\ast}$ that minimizes the quantum
relative entropy $S(\rho\parallel\rho^{\ast})$ is
$$
\rho^{\ast}=\sum_{n}a_{n,n}|\phi_n,\psi_n\rangle\langle\phi_n,\psi_n|,
\eqno{(3)}
$$
where, $|\phi_n\rangle$ and $|\psi_n\rangle$ are orthogonal states
of each subsystem. The states in Eq.(1) are also called maximally
correlated states and are known to have some interesting
properties. For example, the PPT distillable entanglement is
exactly the same as the relative entropy of entanglement [9].\\
\hspace*{8mm}Now, we consider the phase damping model. The density
matrix satisfies the following master equation in the interaction
picture
$$
\frac{d}{dt}\rho(t)=(L_1+L_2)\rho(t),
\eqno{(4)}
$$
with
$$
L_i\rho=\frac{\gamma_i}{2}[2a^{\dagger}_{i}a_i{\rho}a^{\dagger}_{i}a_i
-(a^{\dagger}_{i}a_i)^2\rho-\rho{(a^{\dagger}_{i}a_i)^2}],
\eqno{(5)}
$$
where, $\gamma_i$($i=1,2$) is the ith mode phase damping
coefficient, and $a^{\dagger}_i$ ($a_i$) is the creation
(annihilation) operator of the ith mode field. For arbitrary
initial states described by the density matrix $\rho_0$, the
solution of Eq.(4) can be obtained,
$$
\rho(t)=\sum^{\infty}_{k_1=0}\sum^{\infty}_{k_2=0}\frac{1}{{k_1}!{k_2}!}
(\gamma_1{t})^{k_1}(\gamma_2{t})^{k_2}\Lambda_{k_1,k_2}(t)\rho_0\Lambda_{k_1,k_2}(t),
\eqno{(6)}
$$
where
$$
\Lambda_{k_1,k_2}(t)=(a^{\dagger}_1a_1)^{k_1}(a^{\dagger}_2a_2)^{k_2}
\exp[-\frac{\gamma_1t}{2}(a^{\dagger}_1a_1)^2-\frac{\gamma_2t}{2}(a^{\dagger}_2a_2)^2].
\eqno{(7)}
$$
If we assume the initial density matrix $\rho_0$ is arbitrary
two-mode continuous variable pure states, i.e.,
$$
\rho_0=\sum_{n,m}\sum_{n^{'},m^{'}}a_{n,m}a^{\ast}_{n^{'},m^{'}}|n,m\rangle\langle{n^{'}},m^{'}|,
\eqno{(8)}
$$
where $|n,m\rangle$ is two-mode particle number state. Then, the
time-evolution density matrix with the initial condition is
calculated as
$$
\rho(t)=\sum_{n,m}\sum_{n^{'},m^{'}}a_{n,m}a^{\ast}_{n^{'},m^{'}}\exp[-\frac{\gamma_1t}{2}
(n-n^{'})^2-\frac{\gamma_2t}{2}(m-m^{'})^2]|n,m\rangle\langle{n^{'}},m^{'}|.
\eqno{(9)}
$$
If the density matrix $\rho(t)$ in Eq.(9) can be expressed as the
similar form of Eq.(1), i.e.,
$$
\rho(t)=\sum_{n_1,n_2}c_{n_1,n_2}(t)|\phi^{'}_{n_1},\psi^{'}_{n_1}\rangle\langle
\phi^{'}_{n_2},\psi^{'}_{n_2}|,
\eqno{(10)}
$$
where, $|\phi^{'}_n\rangle$ and $|\psi^{'}_n\rangle$ are
orthogonal states of modes 1 and 2, the relative entropy of
entanglement of $\rho(t)$ in Eq.(10) can be expressed as,
$$
E_R(\rho(t))=-\sum_{n}c_{n,n}(t)\log_{2}c_{n,n}(t)+\textrm{Tr}(\rho(t)\log_2\rho(t)),
\eqno{(11)}
$$
and the separate state $\rho^{\ast}$ that minimizes the quantum
relative entropy is
$$
\rho^{\ast}(t)=\sum_{n}c_{n,n}(t)|\phi^{'}_n,\psi^{'}_n\rangle\langle\phi^{'}_n,\psi^{'}_n|.
\eqno{(12)}
$$
In what follows, we investigate the relative entropy of
entanglement of pair cat states in phase damping channel. Firstly,
we will briefly outline the definition of pair cat states and the
closely related pair coherent states. For two independent boson
annihilation operators $\hat{a}_1$, $\hat{a}_2$, a pair coherent
state $|\xi,q\rangle$ is defined as an eigenstate of both the pair
annihilation operator $\hat{a}_1\hat{a}_2$ and the number
difference operator
$\hat{Q}=\hat{a}^{\dagger}_1\hat{a}_1-\hat{a}^{\dagger}_2\hat{a}_2$
[16], i.e.,
$$
\hat{a}_1\hat{a}_2|\xi,q\rangle=\xi|\xi,q\rangle,~~~\hat{Q}|\xi,q\rangle=q|\xi,q\rangle,
\eqno{(13)}
$$
where $\xi$ is a complex number and $q$ is a fixed integer.
Without loss of generality, we may set $q\geq0$ and the pair
coherent states can be explicitly expanded as a superposition of
the two-mode Fock states, i.e.,
$$
|\xi,q\rangle=N_q\sum^{\infty}_{n=0}\frac{\xi^n}{\sqrt{n!(n+q)!}}|n+q,n\rangle,
\eqno{(14)}
$$
where $N_q=[|\xi|^{-q}I_{q}(2|\xi|)]^{-1/2}$ is the normalization
constant and $I_q$ is the modified Bessel function of the first
kind of order $q$. It has been suggested by Reid and Krippner that
the non-degenerate parametric oscillator transiently generates
pair coherent states, in the limit of very large parametric
nonlinearity and high-Q cavities[17]. Recently, Munro \textit{et
al.} have shown that the pair coherent states can be used to
improve the detection sensitivity of weak forces[18]. Pair cat
states $|\xi,q,\phi\rangle$ are proposed by Gerry and Grobe [10],
which are defined as superposition of two different pair coherent
states, i.e.,
$$
|\xi,q,\phi\rangle=N_{\phi}[|\xi,q\rangle+e^{i\phi}|-\xi,q\rangle],
\eqno{(15)}
$$
where the normalization constant $N_{\phi}$ is given by
$$
N_{\phi}=\frac{1}{\sqrt{2}}[1+N^2_q\cos\phi\sum^{\infty}_{n=0}\frac{(-1)^n|\xi|^{2n}}
{n!(n+q)!}]^{-\frac{1}{2}}. \eqno{(16)}
$$
It is easy to verify that the states $|\xi,q,\phi\rangle$ are
eigenstates of the operator $(\hat{a}_1\hat{a}_2)^2$ with
eigenvalue $\xi^2$. Gou et al. have proposed a scheme for
generating the pair cat state of motion in a two-dimensional ion
trap [12]. In their scheme, the trapped ion is excited
bichromatically by five laser beams along different directions in
the X-Y plane of the ion trap. Four of these have the same
frequency and can be derived from the same source, reducing the
demands on the experimentalist. It is shown that if the initial
vibrational state is given by a two-mode Fock state, pair cat
states are realized when the system reaches its steady state. Our
following calculation show that pair cat states hold controllable
entanglement. So, it is reasonable to regard the controlled
two-dimensional trapped ion as a reliable source of entanglement.
Recent achievements concerning the transfer of entangled state
have provided us a possible way to map the pair cat state of the
motional freedom of two dimensional trapped ions into freely
propagating optical fields [13]. When the free photon propagates
in the optical fibre, one of the encountered decoherence
mechanisms is the phase damping. In the following, we discuss the
entanglement of pair cat states in phase damping channel. We
assume that the initial state is prepared in pair cat states
$|\xi,q,\phi\rangle$. By making use of Eqs.(6) and (7), we obtain
$$
\rho(t)=N^2_{\phi}N^{2}_q\sum^{\infty}_{n=0}\sum^{\infty}_{m=0}\frac{\exp[-\frac{\gamma_1
+\gamma_2}{2}t(n-m)^2]\xi^n\xi^{\ast{m}}(1+(-1)^ne^{i\phi})(1+(-1)^me^{-i\phi})}
{\sqrt{n!m!(n+q)!(m+q)!}}|n+q,n\rangle\langle{m+q},m|. \eqno{(17)}
$$
The relative entropy of entanglement for $\rho(t)$ is calculated
as
$$
E(t,\xi)=-\sum_nN^2_{\phi}N^2_q\frac{|1+(-1)^ne^{i\phi}|^2|\xi|^{2n}}{n!(n+q)!}
\log_2N^2_{\phi}N^2_q\frac{|1+(-1)^ne^{i\phi}|^2|\xi|^{2n}}{n!(n+q)!}
$$
$$
~~~+\textrm{Tr}(\rho(t)\log_2\rho(t)). \eqno{(18)}
$$
\begin{figure}
\centering
\includegraphics{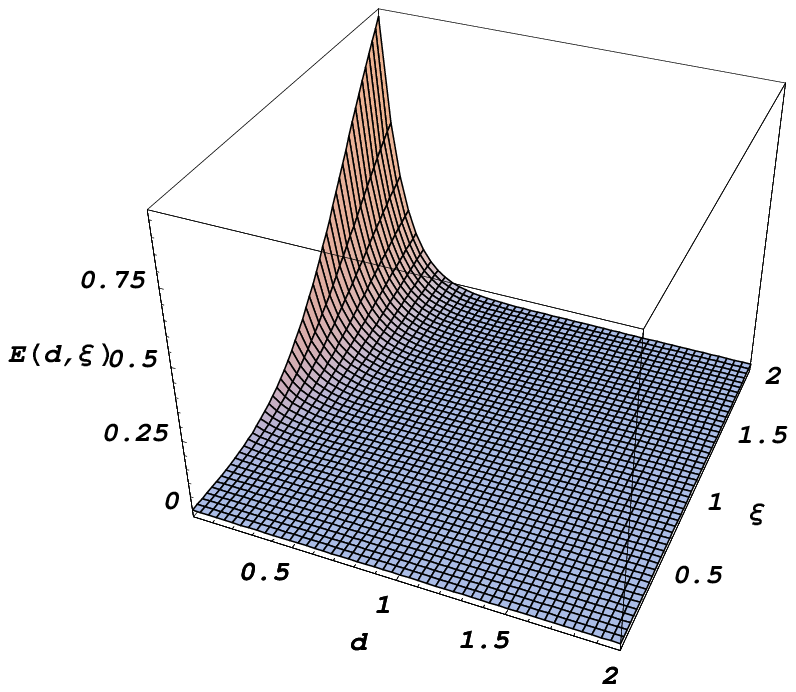}
\caption{The relative entropy of entanglement $E$ of the pair cat
state as a function of the parameter $|\xi|$ and the degree of
damping $d$ for $q=0$ with $\phi=\pi$. \label{Fig.1}}
\end{figure}
\begin{figure}
\centering
\includegraphics{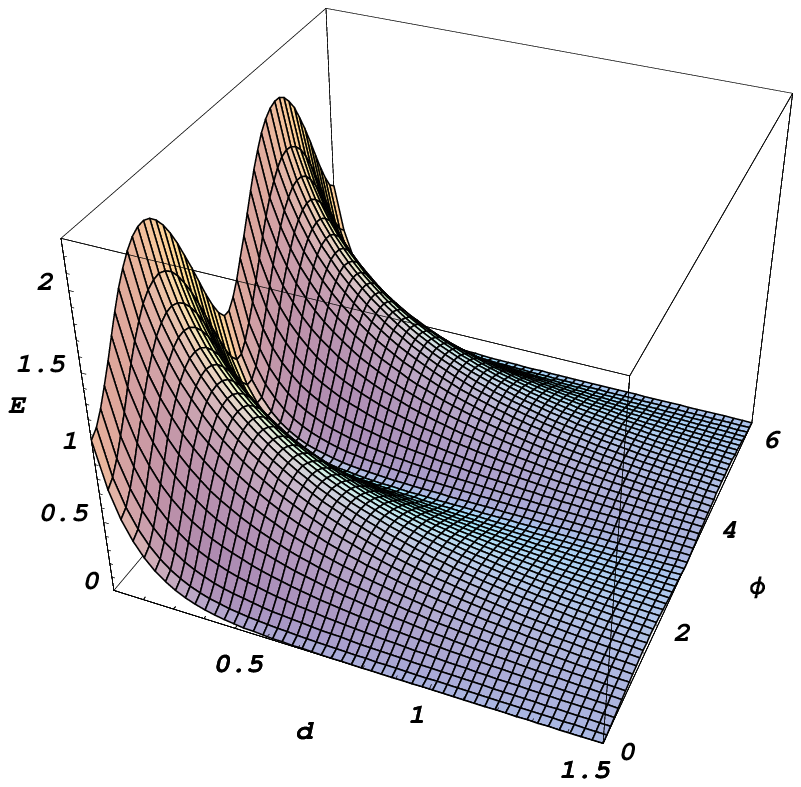}
\caption{The relative entropy of entanglement $E$ of the pair cat
state as a function of the degree of damping $d$ and the parameter
$\phi$  for $q=0$ with $|\xi|=2$. \label{Fig.2}}
\end{figure}
\begin{figure}
\centering
\includegraphics{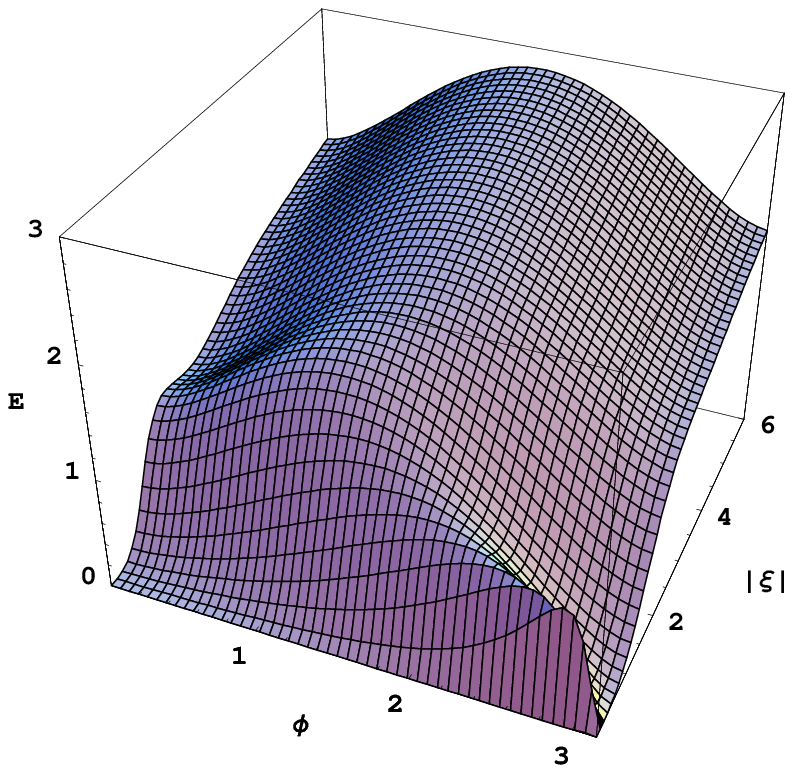}
\caption{The relative entropy of entanglement $E$ of the pair cat
state as a function of the parameter $|\xi|$ and the parameter
$\phi$  for $q=0$ with $d=0$. \label{Fig.3}}
\end{figure}
\begin{figure}
\centering
\includegraphics{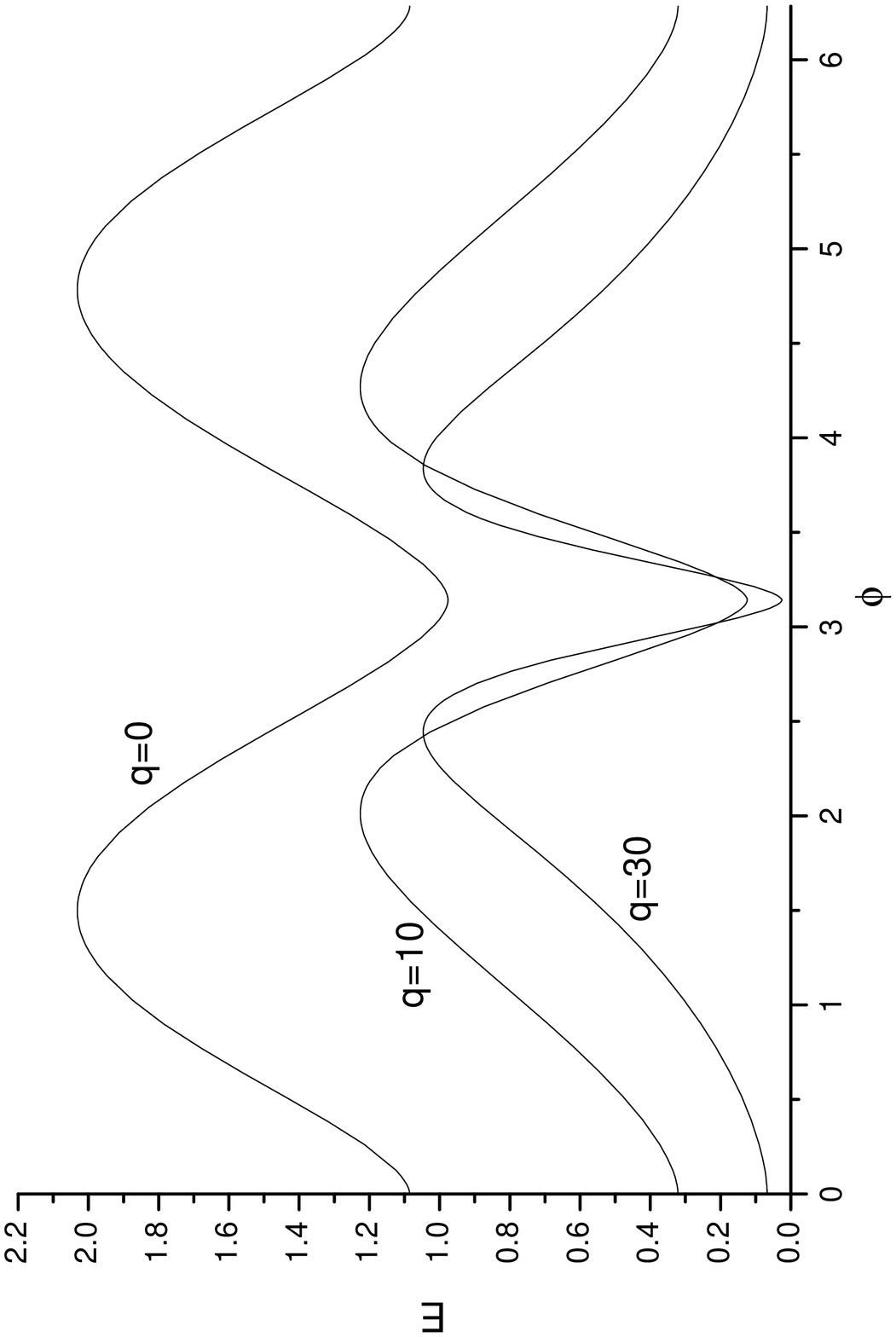}
\caption{The relative entropy of entanglement $E$ of the pair cat
state as a function of the parameter $\phi$ for three values of
$q=0,10$ and $30$ with $d=0$ and $|\xi|=2$. \label{Fig.4}}
\end{figure}
In numerical computations throughout this paper, the parameters
$\gamma_1=\gamma_2=\gamma$, $d=\gamma{t}$ are chosen and the
truncated photon number has been taken to be
$\max(n)=\max(m)=100$, the value of which is sufficiently large
for numerical convergence. Figures 1,2 and 3 show that the
relative entropy of entanglement $E$ of the pair cat state
increases with $|\xi|$ and decreases with degree of damping $d$,
and can be controlled by adjusting the relative phase $\phi$. This
results can be explained as follows: the entanglement of pair cat
states heavily depend on the photon number distribution which can
be modified by the relative phase via the interference. Similar
results have been obtained in Ref.[19]. In Fig.4, we plot the
relative entropy of entanglement $E$ of the pair cat state as a
function of the relative phase $\phi$ for three values of the
parameter $q$. Recently, Hiroshima has numerically calculated the
relative entropy of entanglement of two-mode squeezed vacuum
states, defined by
$|\psi(r)\rangle=\exp[-r(a^{\dagger}_1a^{\dagger}_2-a_1a_2)]|vac\rangle$,
in phase damping channel [20]. It has been shown [21] that the
two-mode squeezed vacuum state in phase damping channel is always
distillable (and inseparable). In the following, we show that the
pair cat states are always distillable (and inseparable) in
phase damping channel.\\
\hspace*{8mm}For two-mode continuous variable states
$|\psi\rangle$,
$$
|\psi\rangle=\sum_nf_n|\phi_n,\psi_n\rangle, \eqno{(19)}
$$
where $|\phi_n\rangle$ and $|\psi_n\rangle$ are orthogonal
particle number states of each subsystem and $f_n$ satisfy the
normalization condition $\sum_n|f_n|^2=1$. The density matrix with
the initial condition $\rho(0)=|\psi\rangle\langle\psi|$ can be
written as
$$
\rho(t)=\sum_{n,m}f_nf^{\ast}_m\exp[-\frac{\gamma_1}{2}t(\phi_n-\phi_m)^2
-\frac{\gamma_2}{2}t(\psi_n-\psi_m)^2]|\phi_n,\psi_n\rangle\langle\phi_m,\psi_m|.
\eqno{(20)}
$$
According to Ref.[22], if operator
$\Omega(t)=\textrm{Tr}_D\rho(t)\otimes{I}-\rho(t)$ is not positive
definite, there is always a scheme to distill $\rho(t)$. Here, we
find
$\textrm{Tr}_D\rho(t)=\sum_n|f_n|^2|\phi_n\rangle\langle\phi_n|$.
If there are two nonzero $f_i$, $f_j$, it is always possible to
choose four vectors
$|W_1\rangle=\frac{1}{\sqrt{2}}(|\phi_i,\psi_i\rangle+|\phi_j,\psi_j\rangle)$,
$|W_2\rangle=\frac{1}{\sqrt{2}}(|\phi_i,\psi_i\rangle-|\phi_j,\psi_j\rangle)$,
$|W_3\rangle=\frac{1}{\sqrt{2}}(|\phi_i,\psi_i\rangle+i|\phi_j,\psi_j\rangle)$,
$|W_4\rangle=\frac{1}{\sqrt{2}}(|\phi_i,\psi_i\rangle-i|\phi_j,\psi_j\rangle)$.
Then, we have
$$
\Omega_1(t)\equiv\langle{W}_1|\Omega(t)|W_1\rangle=-\exp[-\frac{\gamma_1}{2}t(\phi_i
-\phi_j)^2-\frac{\gamma_2}{2}t(\psi_i-\psi_j)^2]\textrm{Re}(f_if^{\ast}_j),
$$
$$
\Omega_2(t)\equiv\langle{W}_2|\Omega(t)|W_2\rangle=\exp[-\frac{\gamma_1}{2}t(\phi_i
-\phi_j)^2-\frac{\gamma_2}{2}t(\psi_i-\psi_j)^2]\textrm{Re}(f_if^{\ast}_j),
$$
$$
\Omega_3(t)\equiv\langle{W}_3|\Omega(t)|W_3\rangle=\exp[-\frac{\gamma_1}{2}t(\phi_i
-\phi_j)^2-\frac{\gamma_2}{2}t(\psi_i-\psi_j)^2]\textrm{Im}(f_if^{\ast}_j),
$$
$$
\Omega_4(t)\equiv\langle{W}_4|\Omega(t)|W_4\rangle=-\exp[-\frac{\gamma_1}{2}t(\phi_i
-\phi_j)^2-\frac{\gamma_2}{2}t(\psi_i-\psi_j)^2]\textrm{Im}(f_if^{\ast}_j),
\eqno{(21)}
$$
which satisfy
$$
\Omega_1(t)+\Omega_2(t)\equiv0,~~~\Omega_3(t)+\Omega_4(t)\equiv0,
$$
$$
\Omega_1(t)+i\Omega_4(t)=-\exp[-\frac{\gamma_1}{2}t(\phi_i-\phi_j)^2
-\frac{\gamma_2}{2}t(\psi_i-\psi_j)^2](f_if^{\ast}_j)\neq0,
$$
$$
\Omega_2(t)+i\Omega_3(t)=\exp[-\frac{\gamma_1}{2}t(\phi_i-\phi_j)^2
-\frac{\gamma_2}{2}t(\psi_i-\psi_j)^2](f_if^{\ast}_j)\neq0,
\eqno{(22)}
$$
Eqs.(22) show that there is at least one of $\Omega_k(t)$
($k=1,2,3,4$), which is negative for any finite
$\frac{\gamma_1+\gamma_2}{2}t$. From the above, we obtain the
following conclusion: the two-mode continuous variable state
$|\psi\rangle=\sum_nf_n|\phi_n,\psi_n\rangle$, in which
$|\phi_n\rangle$ and $|\psi_n\rangle$ are orthogonal particle
number states of each subsystem, is always distillable (and
inseparable) in phase damping channel, if there are at least two
nonzero values of coefficiences $f_n$. Obviously, pair coherent
states and pair cat states which belong to the family of states in
Eq.(20) is always distillable in phase damping channel. It should
be interesting to consider a slightly modified purification
protocol similar to the protocol in Ref.[14] to distill maximal
entangled states from the mixed pair cat states or mixed pair
coherent states due to phase damping.\\

\section * {III. FIDELITY OF TELEPORTATION VIA PAIR CAT STATES IN PHASE DAMPING CHANNEL}
\hspace*{8mm}Recently, Cochrane et al. have presented a
teleportation protocol by making use of joint measurements of the
photon number sum and phase difference on two field modes [11].
Various kinds of two modes entangled states used as the
entanglement resource have been discussed and the respective
teleportation fidelities have been investigated. In this section,
we adopt the protocol of cochrane et al. to investigate the
fidelity of teleportation, in which the pair cat state is utilized
as the entanglement resource. The influence of phase damping
on the fidelity is also discussed.\\
\hspace*{8mm}Consider arbitrary target state sent by Alice to Bob
$$
|\psi\rangle_T=\sum^{\infty}_{k=0}d_k|k\rangle_T, \eqno{(23)}
$$
where $|k\rangle_T$ is the fock state. Initially, Alice and Bob
share the two-mode fields in the pair cat state. Then, the total
state is
$$
|\psi\rangle=N_{\phi}N_{q}\sum^{\infty}_{n=0}\sum^{\infty}_{k=0}
\frac{d_k\xi^n[1+(-1)^ne^{i\phi}]}{\sqrt{n!(n+q)!}}
|n+q\rangle_a|n\rangle_b|k\rangle_T. \eqno{(24)}
$$
The whole operation of this teleportation protocol can be
decomposed as two steps: Alice makes a joint measurement of the
photon number sum and phase difference of the target state and her
component of the pair cat state; The results of the joint
measurement are sent to Bob via the classical channel, and Bob
reproduce the target state after appropriate amplification and
phase shift operations according to the results of the joint
measurement. The joint measurement of the photon number sum and
phase difference has attracted much attention due to its extensive
potential applications both in quantum optics and quantum
information [23,24]. In Ref.[24], Luis et al. introduced the
hermitian phase-difference operator
$$
\hat{\Theta}_{12}=\sum^{\infty}_{N=0}\sum^{N}_{r=0}\theta^{N}_r|\theta^N_r
\rangle\langle\theta^N_r|,
\eqno{(25)}
$$
with
$$
|\theta^N_r\rangle=\frac{1}{\sqrt{N+1}}\sum^{N}_{n=0}e^{in\theta^N_r}|n\rangle_1|N-n\rangle_2,
\eqno{(26)}
$$
and
$$
\theta^N_r=\vartheta+\frac{2\pi{r}}{N+1},
\eqno{(27)}
$$
where $\vartheta$ is an arbitrary angle. It is obvious that the
joint measurement projects the two-mode quantum state onto
$|\theta^N_r\rangle$. In Ref.[25], a physical scheme of the joint
measurement of the photon number sum and phase difference of
two-mode fields was proposed, in which only the linear optical
elements and single-photon detector are involved.\\
\hspace*{8mm}If Alice measure the number sum $\hat{N}_a+\hat{N}_T$
of the target and her component of the pair cat state with result
$N$, the state of the total system is projected onto
$$
|\psi^{(N)}\rangle=[\frac{P(\xi,q,\phi,N)}{N^2_qN^2_{\phi}}]^{-1/2}\sum^{N}_{l=q}
\frac{d_{N-l}\xi^{l-q}[1+(-1)^{l-q}e^{i\phi}]}
{\sqrt{l!(l-q)!}}|l\rangle_a|l-q\rangle_b|N-l\rangle_T,
\eqno{(28)}
$$
where
$$
P(\xi,q,\phi,N)=N^2_{\phi}N^2_{q}\sum^{N}_{l=q}\frac{2|d_{N-l}
\xi^{l-q}|^2[1+(-1)^{l-q}\cos\phi]}{l!(l-q)!},
\eqno{(29)}
$$
is the probability of obtaining the result $N$. Further
measurement of phase difference with the result $\theta_{-}$
performed by Alice will project Bob's mode onto the pure state
$$
|\psi^{(N,\theta_{-})}\rangle=[\frac{P(\xi,q,\phi,N)}{N^2_qN^2_{\phi}}]^{-1/2}
\sum^{N-q}_{n=0}\frac{d_{N-q-n}(e^{-i\theta_{-}}\xi)^n[1+(-1)^ne^{i\phi}]}
{\sqrt{n!(n+q)!}}|n\rangle_b.
\eqno{(30)}
$$
Alice sends the values $N$ and $\theta_{-}$ to Bob, and then Bob
amplifies his mode so that $|n\rangle_b\rightarrow|N-q-n\rangle_b$
[26] and makes a operation $e^{-i\hat{N}_b\theta_{-}}$ for phase
shifting his mode. The teleportation protocol is then completed
and Bob finally has the state in
$$
|\psi^{(N)}\rangle_{b}=[\frac{P(\xi,q,\phi,N)}{N^2_qN^2_{\phi}}]^{-1/2}
\sum^{N-q}_{n=0}\frac{d_{N-q-n}\xi^n[1+(-1)^ne^{i\phi}]}{\sqrt{n!(n+q)!}}|N-q-n\rangle_b.
\eqno{(31)}
$$
The fidelity of this protocol depends on the result $N$ and can be
obtained as follows
$$
F(\xi,q,\phi,N)=[\frac{P(\xi,q,\phi,N)}{N^2_qN^2_{\phi}}]^{-1}|\sum^{N-q}_{n=0}
\frac{|d_{N-q-n}|^2\xi^n[1+(-1)^ne^{i\phi}]}{\sqrt{n!(n+q)!}}|^2.
\eqno{(32)}
$$
The average fidelity defined by
$\bar{F}(\xi,q,\phi)\equiv\sum^{\infty}_{N=q}P(\xi,q,\phi,N)F(\xi,q,\phi,N)$
is
$$
\bar{F}(\xi,q,\phi)=N^2_qN^2_{\phi}\sum^{\infty}_{N=0}|\sum^{N}_{n=0}
\frac{|d_{N-n}|^2\xi^n[1+(-1)^ne^{i\phi}]}{\sqrt{n!(n+q)!}}|^2.
\eqno{(33)}
$$
Let the target state be a coherent state
$|\psi\rangle_T=|\alpha\rangle_T$. Then, the average fidelity can
be expressed as
$$
\bar{F}(\xi,q,\phi,\alpha)=N^2_qN^2_{\phi}e^{-2|\alpha|^2}\sum^{\infty}_{N=0}|
\alpha|^{4N}|\sum^{N}_{n=0}\frac{|\alpha|^{-2n}\xi^n[1+(-1)^ne^{i\phi}]}{(N-n)!
\sqrt{n!(n+q)!}}|^2.
\eqno{(34)}
$$
\begin{figure}
\centering
\includegraphics{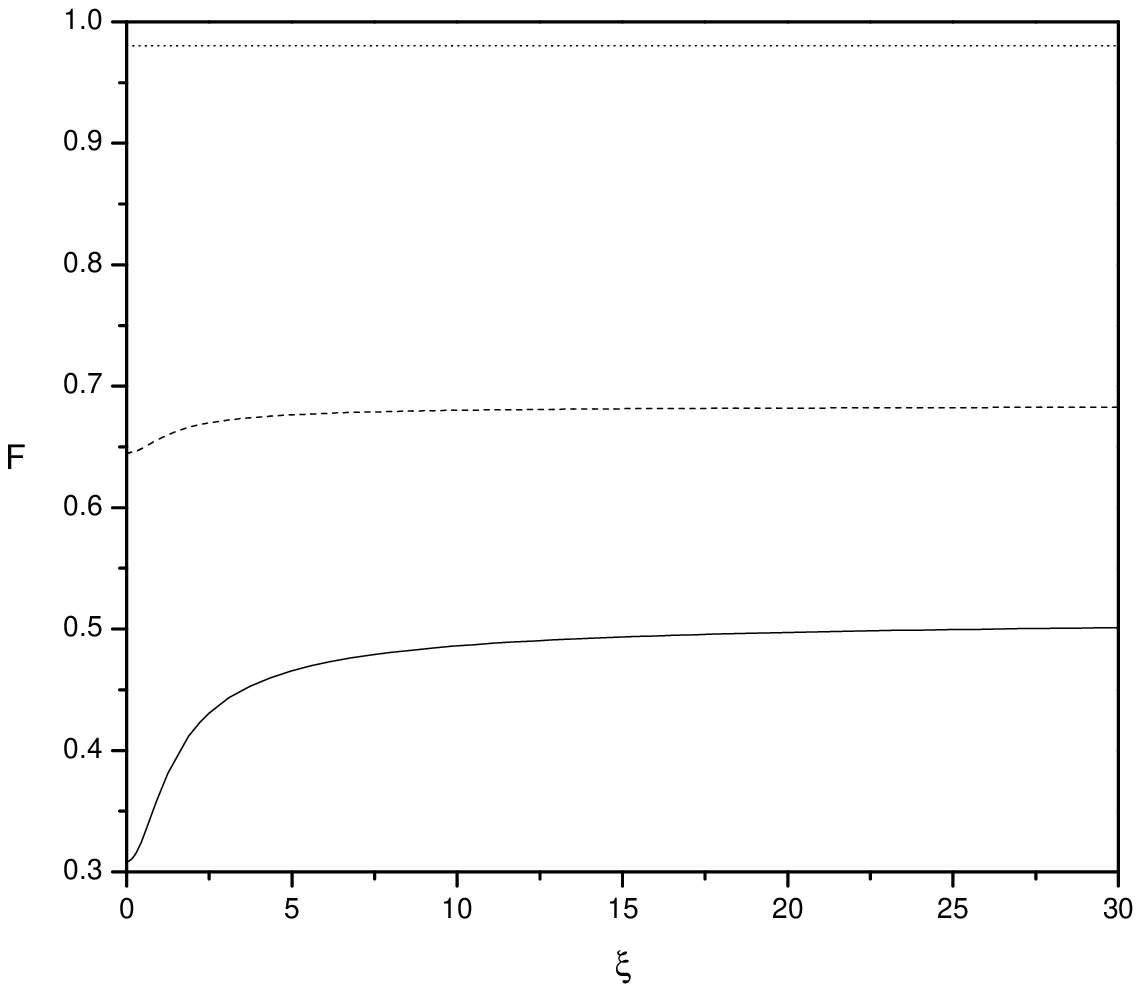}
\caption{The average fidelity is plotted as the functions of the
parameters $\xi$ with $q=0$ and $\phi=\pi/2$ for different values
of $|\alpha|$, $|\alpha|=1$ (Solid Line), $|\alpha|=0.5$ (Dash
Line), $|\alpha|=0.1$ (Dot Line). \label{Fig.5}}
\end{figure}
In Fig.5, we have plotted the average fidelity as the functions of
the parameters $\xi$ for different values of $|\alpha|$. It is
shown that the average fidelity increases with the value of $\xi$.
Furthermore, the average fidelity defined above heavily depends on
the teleported states. If the teleported state is a coherent
state, the smaller the amplitude of coherent states, the higher
the average fidelity. Its physical reason can be elucidated by two
facts: one fact is that this protocol works perfectly if the
target is a number state [11]; the other fact is that the smaller
the amplitude of a coherent state, the closer the state distance
between the coherent state and a specific number state,i.e., the
vacuum state. In what follows, we discuss the influence of phase
damping on the fidelity of the above teleportation protocol. In
this case, the state of the total system can be written as follows
$$
\rho_T=\sum^{\infty}_{k,l,n,m=0}
\frac{d_kd^{\ast}_l\xi^n\xi^{\ast{m}}[1+(-1)^ne^{i\phi}][1+(-1)^me^{-i\phi}]
e^{-\gamma{t}(n-m)^2}}{\sqrt{n!m!(n+q)!(m+q)!}}|k\rangle_{TT}\langle{l}|
\otimes|n+q,n\rangle\langle{m+q},m|.
\eqno{(35)}
$$
After completing the protocol described above, Bob finally
achieves the state in his mode expressed by
$$
\rho_{b}=[\frac{P(\xi,q,\phi,N)}{N^2_qN^2_{\phi}}]^{-1}\sum^{N^{\prime}}_{n,m=0}
\frac{d_{N^{\prime}-n}d^{\ast}_{N^{\prime}-m}\xi^n\xi^{\ast{m}}[1+(-1)^ne^{i\phi}][1+(-1)^me^{-i\phi}]
e^{-\gamma{t}(n-m)^2}}{\sqrt{n!m!(n+q)!(m+q)!}}|N^{\prime}-n\rangle_{bb}\langle{N^{\prime}-m}|,
\eqno{(36)}
$$
where $N^{\prime}=N-q$ and $N$ is the measured number sum of
Alice's joint measurement. Then, the fidelity of this protocol
depends on the result $N$ and is
$$
F(\xi,q,\phi,N,\gamma{t})=[\frac{P(\xi,q,\phi,N)}{N^2_qN^2_{\phi}}]^{-1}\sum^{N^{\prime}}_{n,m=0}
\frac{|d_{N^{\prime}-n}|^2|d_{N^{\prime}-m}|^2\xi^n\xi^{\ast{m}}[1+(-1)^ne^{i\phi}][1+(-1)^me^{-i\phi}]
e^{-\gamma{t}(n-m)^2}}{\sqrt{n!m!(n+q)!(m+q)!}}.
\eqno{(37)}
$$
The average fidelity is
$$
\bar{F}(\xi,q,\phi,\gamma{t})=N^2_qN^2_{\phi}\sum^{\infty}_{N^{\prime}=0}\sum^{N^{\prime}}_{n,m=0}
\frac{|d_{N^{\prime}-n}|^2|d_{N^{\prime}-m}|^2\xi^n\xi^{\ast{m}}[1+(-1)^ne^{i\phi}][1+(-1)^me^{-i\phi}]
e^{-\gamma{t}(n-m)^2}}{\sqrt{n!m!(n+q)!(m+q)!}}.
\eqno{(38)}
$$
For a coherent state $|\alpha\rangle$ teleported by Alice, the
average fidelity can be rewritten as
$$
\bar{F}(\xi,q,\phi,|\alpha|,\gamma{t})=N^2_qN^2_{\phi}e^{-2|\alpha|^2}\sum^{\infty}_{N^{\prime}=0}|\alpha|^{4N^{\prime}}\sum^{N^{\prime}}_{n,m=0}
\frac{|\alpha|^{-2n-2m}\xi^n\xi^{\ast{m}}[1+(-1)^ne^{i\phi}][1+(-1)^me^{-i\phi}]
e^{-\gamma{t}(n-m)^2}}{(N^{\prime}-n)!(N^{\prime}-m)!\sqrt{n!m!(n+q)!(m+q)!}}.
\eqno{(39)}
$$
In Fig.6, the average fidelity for the coherent states is plotted
as a function of $\gamma{t}$ for different values of amplitude
$|\alpha|$. It is shown that the phase damping deteriorate the
average fidelity of teleportation basing on the pair cat state,
which is qualitatively consistent with the behavior of its
relative entropy of entanglement in the phase damping channel. As
mentioned above, the essential parts of this teleportation
protocol are the preparation of the pair cat state and the joint
measurement of the number sum and phase difference. The direct
preparation of pair cat states in two modes optical fields is
still a open question. However, we can map the pair cat state of
the motional freedom of two dimensional trapped ions into freely
propagating optical fields. The details will be discussed
elsewhere. Recently, the physical implementation of joint
measurement of photon number sum and phase difference of two-mode
optical fields is shown to be possible by using only the linear
optical elements and the single-photon detector [25]. So we can
conclude that the physical realization of this teleportation
protocol is feasible at the present technology.

\section * {IV. CONCLUSION}
\hspace*{8mm}In this paper, we investigate the entanglement of
pair cat states in the phase damping channel by employing the
relative entropy of entanglement. We give the numerical
calculations of the relative entropy of entanglement of this state
\begin{figure}
\centering
\includegraphics{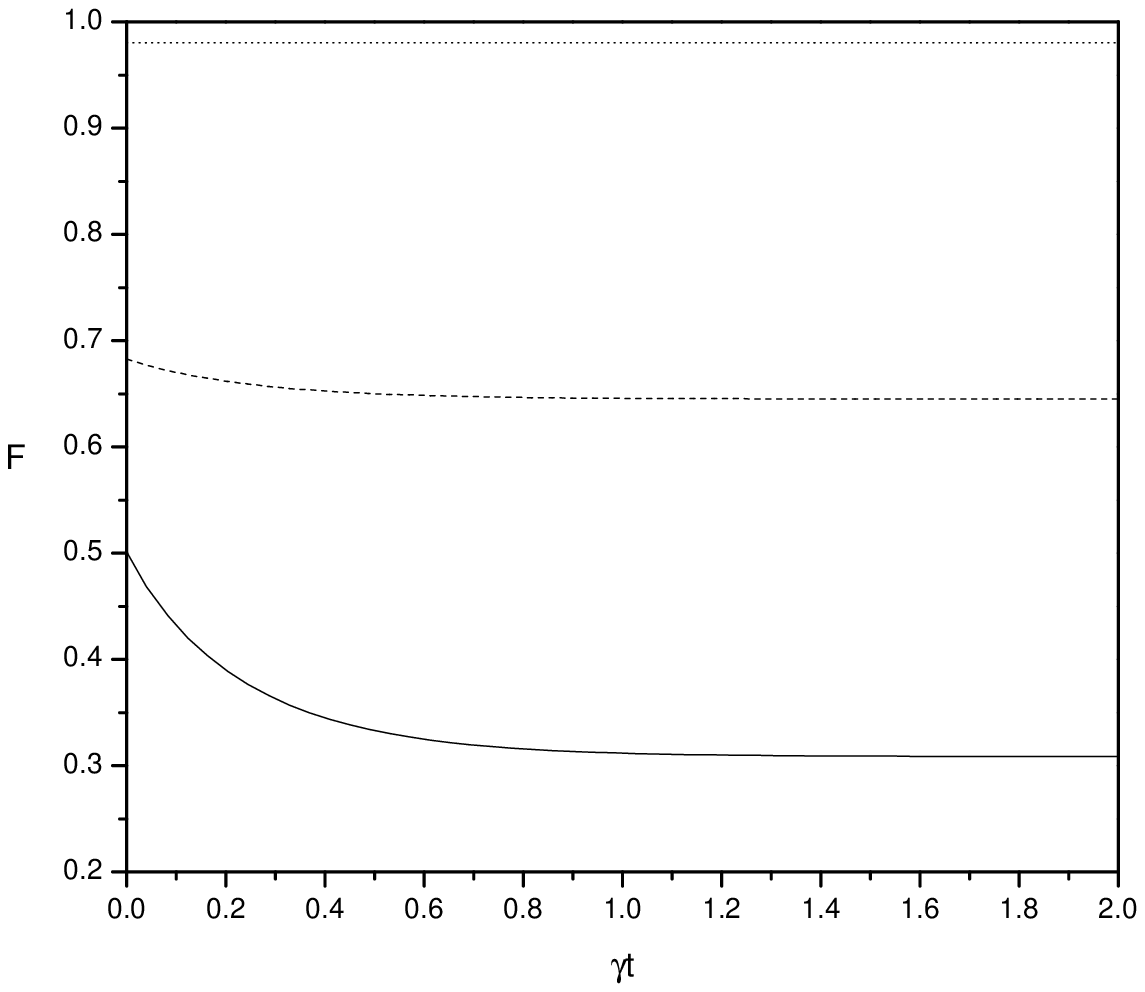}
\caption{The average fidelity is plotted as the functions of the
phase damping coefficience $\gamma{t}$ with $\xi$=30, $q=0$ and
$\phi=\pi/2$ for different values of $|\alpha|$, $|\alpha|=1$
(Solid Line), $|\alpha|=0.5$ (Dash Line), $|\alpha|=0.1$ (Dot
Line). \label{Fig.6}}
\end{figure}
in the phase damping channel and study the influence of the
parameters on the relative entropy of entanglement. We find that
the relative phase of the pair cat state can control the relative
entropy of entanglement. Then, we show that the pair cat states
can always be distillable in the phase damping channel. Finally,
we analyze the fidelity of teleportation for the pair cat states
by using joint measurements of the photon-number sum and phase
difference. The influence of phase damping on the fidelity is
discussed. The behavior of average fidelity for teleporting a
coherent based on the pair cat state is qualitatively consistent
with its relative entropy of entanglement. It is interesting to
investigate the entanglement and teleportation fidelity of pair
cat states in the amplitude damping channel.
\section * {ACNOWLEDGMENTS}
This project was supported by the National Natural Science
Foundation of China (Project NO.10174066).

\end{document}